\documentclass[3p,times]{elsarticle}

\usepackage{ecrc}
\usepackage{subfigure}

\volume{00}

\firstpage{1}

\journalname{Nuclear Physics A}

\runauth{Marta Verweij for the ALICE collaboration}

\jid{nupha}

\usepackage{amssymb}

\usepackage[figuresright]{rotating}

\newcommand{\sNN}{\ensuremath{\sqrt{s_{\mathrm{NN}}}}}
\newcommand{\rom}[1]{{\mathrm{#1}}}   
\newcommand{\PbPb}{Pb--Pb }
\newcommand{\pp}{pp }
\newcommand{\pt}{\ensuremath{p_\rom{T}}}
\newcommand{\deltapt}{\ensuremath{\rom{\delta}\!\pt}}

\newcommand{\GeVc}{GeV/\ensuremath{c}}
\newcommand{\kt}{\ensuremath{k_{\rom{T}}}}

\newcommand{\Fig}[4]{
    \begin{figure}[!h]
        \centering
        \includegraphics[#3]{#1}
        \caption{\label{#2}\small#4}
    \end{figure}
}

\newcommand{\SubFig}[9]{
  \begin{figure}[!h]
    \centering
    \subfigure[#3]{\label{#2}
      \includegraphics[#9]{#1}}
    \hspace{.1in}
    \subfigure[#6]{\label{#5}
      \includegraphics[#9]{#4}}
    \caption{#7\label{#8}}
  \end{figure}
}

\newlength{\myfigwidth}
\newlength{\mydoublefigwidth}
\newlength{\figurematrixwidth}
\begin{document}

\begin{frontmatter}



\dochead{}

\title{Measurement of jet spectra in Pb-Pb collisions at
$\sqrt{s_{NN}}$=2.76 TeV with the ALICE detector at the LHC}


\author{Marta Verweij\fnref{fn1} (for the ALICE collaboration)}
\address[fn1]{Utrecht University}
\ead{marta.verweij@cern.ch}

\address{}

\begin{abstract}
We report a measurement of transverse momentum spectra of jets detected with
the ALICE detector in Pb-Pb collisions at $\sNN$=$2.76$ TeV. Jets are
reconstructed from charged particles using the anti-\kt\ jet algorithm. The
background from soft particle production is determined for each event and
subtracted. The remaining influence of underlying event fluctuations is
quantified by embedding different probes into heavy-ion data. The
reconstructed transverse momentum spectrum is corrected for background
fluctuations by unfolding. We compare the inclusive jet spectra reconstructed
with $R=0.2$ and $R=0.3$ for different centrality classes and compare the jet
yield in \PbPb and pp events.
\end{abstract}


\end{frontmatter}


\section{Jet Reconstruction with charged particles in
  ALICE}\label{sec:JetReco} 
For this analysis data collected by the ALICE experiment in the heavy-ion run
of the LHC in the fall of $2010$ with an energy $\sNN = 2.76$ TeV are used.
Jets are clustered from charged particles reconstructed using the central
tracking detectors Inner Tracking System (ITS) and Time Projection Chamber
(TPC). This ensures a uniform acceptance in full azimuth and $|\eta|<0.9$.

For signal jets the anti-\kt\ algorithm \cite{Cacciari2008b} is used and for
background clusters the \kt\ algorithm \cite{Cacciari2006}. 
For this analysis jet radii of $R=0.2$ and $0.3$ were used. The minimum \pt\ of the
jet constituents is $0.15$ \GeVc. All jets with a jet axis within
$|\eta|<0.5$ are considered for this analysis.
The average background density per unit area $\rho$ is estimated
event-by-event by calculating the median $\pt /A$ (with $A$ the area of the
jet) of all except the two leading \kt\ clusters in the event. From each
signal anti-\kt\ jet in the event $\rho \cdot A$ is subtracted from the
reconstructed \pt\ of the jet \cite{Abelev:2012ej, Cacciari2010}.

\section{Unfolding}\label{sec:Unfolding}
ALICE has measured background fluctuations induced by the underlying event in
heavy-ion collisions and studied the contributing sources
\cite{Abelev:2012ej}. Background fluctuations have a large impact on the
measured jet spectrum due to the tail to upwards fluctuations in the
\deltapt\ distribution. The fluctuations are corrected for by unfolding. A
response matrix $RM_{\deltapt}$ containing the \pt\ smearing due to background
fluctuations is constructed from the measured \deltapt\ distribution.

Detector effects affecting the jet energy resolution are the charged particle
tracking efficiency and the transverse momentum resolution. The efficiency is
the dominant contributor to the jet energy resolution. The tracking efficiency has
been studied with the detector simulation using Pythia and HIJING. The magnitude of
the correction of detector effects is a $10$\% shift in the jet energy scale
which corresponds to $40$\% on the jet yield assuming the jet spectrum scales with $p_{T}^{-5}$. 
The uncertainty on the exact knowledge of the tracking efficiency results in a $3$\% uncertainty on
the jet energy scale reported in the systematic uncertainty of the
measurement.

The two response matrices from background fluctuations and detector effects
are combined to obtain the response matrix which will be used in the unfolding
procedure: $M = RM_{\deltapt} \cdot RM_{det} \cdot T$ in which $M$ is the
measured jet yield and $T$ is the true jet yield.

Jet spectra are unfolded using a $\chi^{2}$ minimization method. Using this
method the number of jets is always conserved. The $\chi^{2}$
function to be minimized indicates how well the unfolded
distribution convoluted with the response matrix (the refolded spectrum)
describes the measured spectrum. The $\chi^{2}$ function used in this analysis
is:
\begin{equation}\label{eq:chi2function}
\chi^{2} =
\displaystyle\sum_{\mathrm{refolded}}\left(\frac{y_{\mathrm{refolded}}-y_{\mathrm{measured}}}{\sigma_{\mathrm{measured}}}\right)^{2}
+ \beta\displaystyle\sum_{\mathrm{unfolded}}\left(\frac{d^{2}log\; y_{\mathrm{unfolded}}}{d\;
  log\; \pt^{2}}\right) ^{2},
\end{equation}
in which $y$ is the yield of the refolded, measured or unfolded jet spectrum and $\sigma_{\rom{measured}}$ the statistical
uncertainty on the measured jet spectrum.
The first summation term of equation \ref{eq:chi2function} gives the
$\chi^{2}$ between the refolded spectrum and the measured jet spectrum. The
second summation term of equation \ref{eq:chi2function} is the penalty term
which is used to regularize the unfolded solution and favors a local power
law. Regularization is necessary to avoid heavily fluctuating solutions. The
strength of the applied regularization $\beta$ is tuned so as to make sure the
regularization term is not dominant. In case the regularization is dominant
the penalty term is equal to or larger than the $\chi^{2}$ between
the refolded and measured spectrum. In this case the refolded spectrum does
not describe the measured spectrum. In case the regularization is too weak or
too strong off-diagonal correlations in the Pearson coefficients extracted
from the covariance matrix are observed.

The regularization adds a systematic uncertainty of $\sim10\%$ for central
events and $\sim4\%$ for peripheral events to the unfolded yield .

\Fig{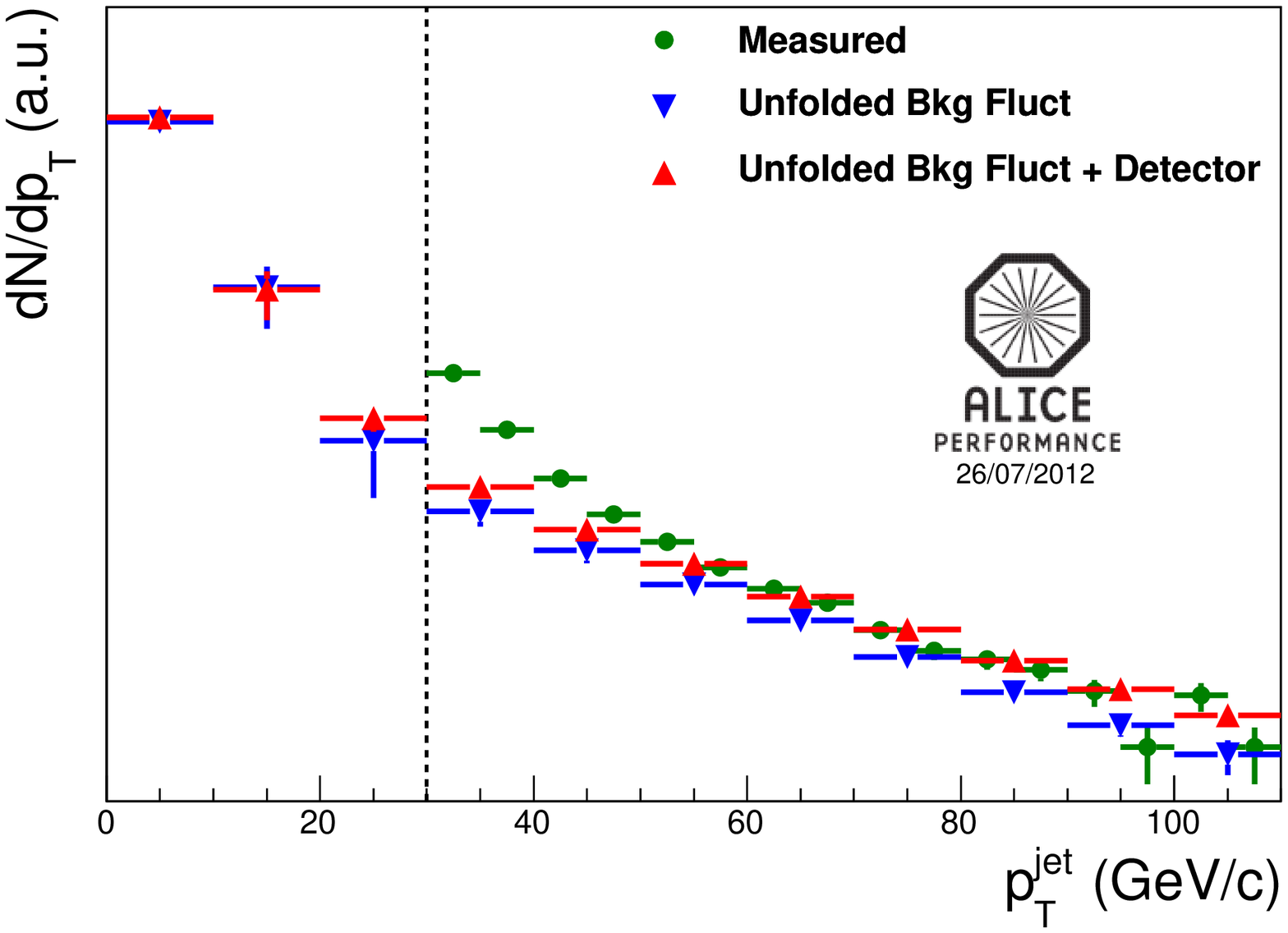}{fig:PerformanceUnfolding}{width=0.4\linewidth}{
  Unfolded spectra after correction for background fluctuations and the
  combined correction for background fluctuations and detector effects are
  shown. The dotted vertical line at $\pt=30$ GeV/$c$ indicates the minimum
  \pt\ cut-off of the measured spectrum.}
{\bf Transverse momentum range to optimize}\newline 
The measured spectrum is only used between a minimum and maximum transverse momentum,
$\pt ^{\rom{min,meas}}$ and $\pt ^{\rom{max,meas}}$. The maximum \pt\ cut-off
is driven by the available statistics. The minimum \pt\ cut-off is introduced
to suppress clusters from the soft background which do not originate from a
hard process. These soft clusters dominate the low \pt\ part of the jet
spectrum. The optimal value of the minimum \pt\ cut-off has been studied in a
model in which a jet spectrum as in vacuum is folded with the measured
background fluctuations and using the jet background model described in
\cite{deBarros:2011ph, GBarrosHP2012}.  The minimum \pt\ cut off on the
measured spectrum is typically $5\sigma(\deltapt)$ in which $\sigma(\deltapt)$
is the width of the \deltapt\ distribution.

{\bf Transverse momentum range of unfolded spectrum}\newline 
In the unfolding it is allowed to redistribute the yield of the measurement below
the minimum \pt\ cut-off on the measured spectrum. Combinatorial
clusters with $\pt>\pt ^{\rom{min,meas}}$ in the measured spectrum will appear in the low \pt\ region of the
unfolded spectrum. The low \pt\ region of the unfolded spectrum also allows
feed-in from $\pt<\pt ^{\rom{min,meas}}$ into the region where the $\chi^{2}$ is
minimized. Feed-in from larger transverse momenta than the maximum
measured momentum is also allowed by extending the reach of the unfolded
spectrum to $\pt = 250$ \GeVc.

The unfolding will break down if the fit gets too much freedom which is when
the number of fitting parameters, number of bins in the unfolded spectrum, are similar or larger than the number of bins in the measurement.

Figure \ref{fig:PerformanceUnfolding} illustrates the different \pt\ ranges for the
measured and unfolded spectrum. In this example there are $16$
bins in the measured spectrum and $11$ in the unfolded. The unfolded spectrum
below $\pt ^{min,meas}$ is not used in the measurement.

\section{Results}
\begin{figure}[!h]
  \centering
  \subfigure[Unfolded jet
  spectra for jet radius $R=0.3$.]{\label{fig:UnfSpectra_R03}
    \includegraphics[width=0.36\linewidth]{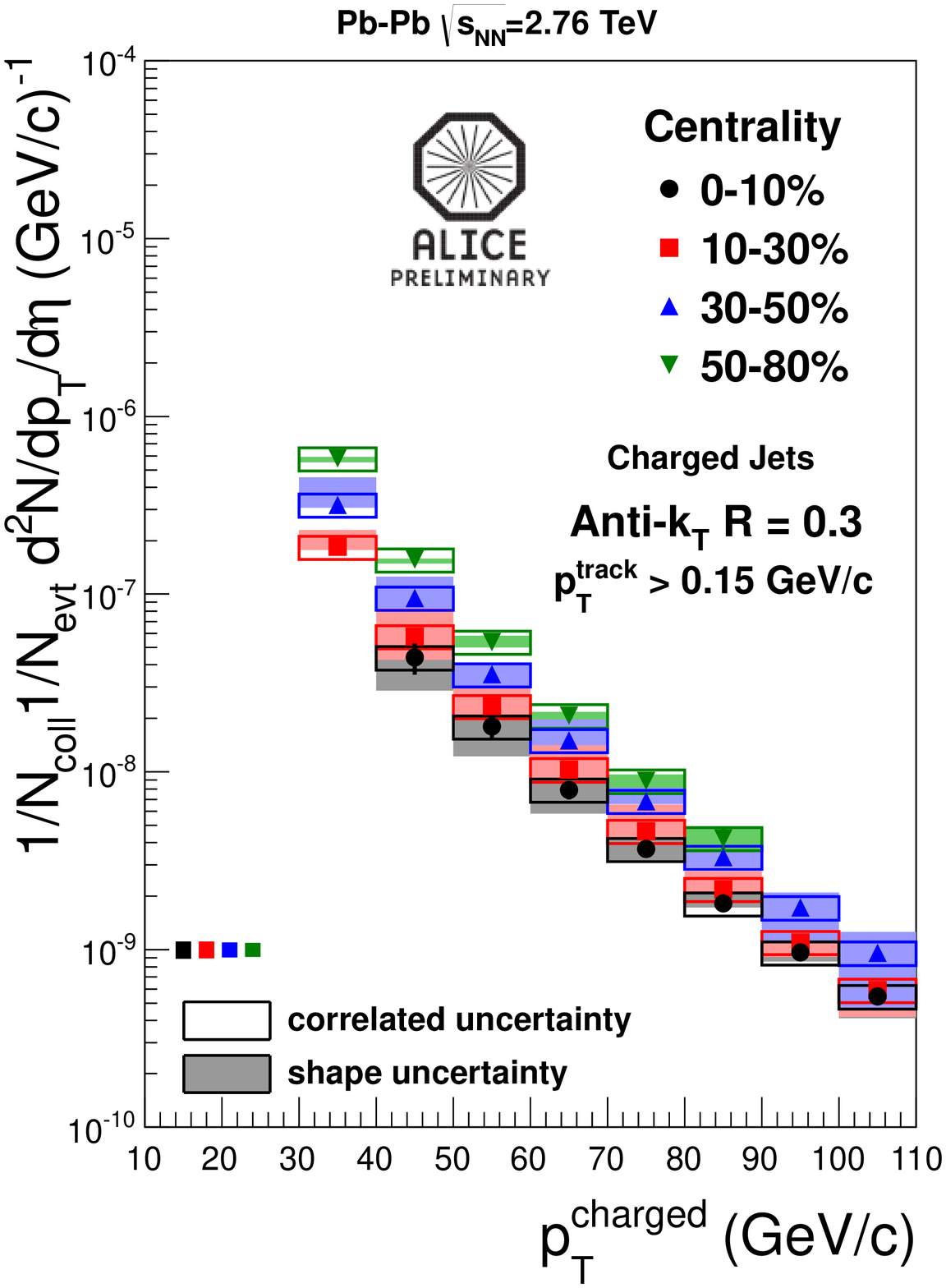}}
  \hspace{.3in}
  \subfigure[$R_{AA}^{Pythia}$ for jet
  radius $R=0.3$. For the reference jet spectrum  Pythia Perugia0 is used.]{\label{fig:RAA_R03}
    \includegraphics[width=0.42\linewidth]{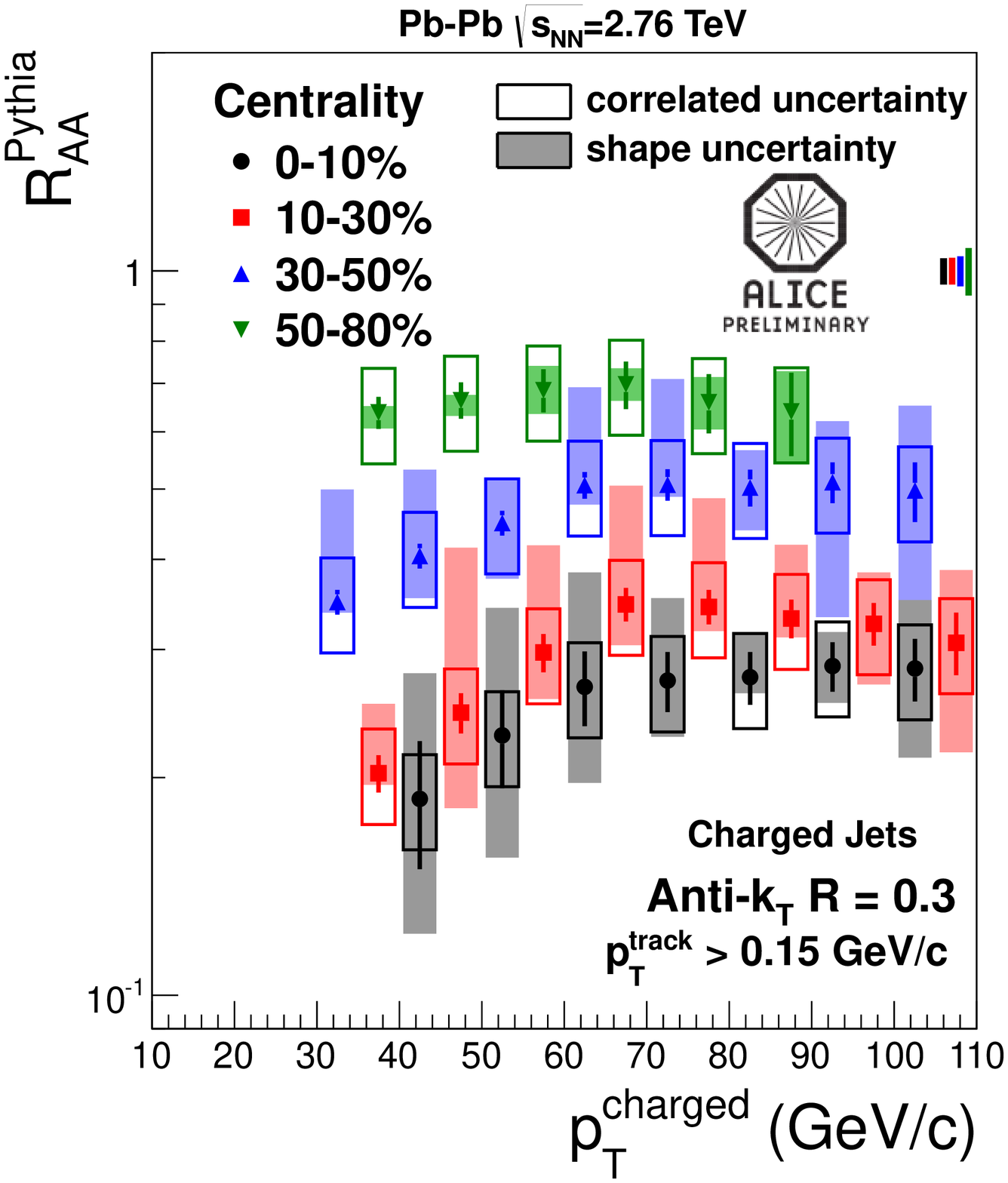}}
  \caption{Unfolded jet spectra and nuclear modification factor $R_{AA}^{Pythia}$ for jets reconstructed with radius
  $R=0.3$.\label{fig:UnfSpectraRAAR03}}
\end{figure}

The corrected differential jet spectrum normalized by the number of collisions
$N_{\mathrm{coll}}$ reconstructed from charged particles in heavy-ion
collisions with jet radius $R=0.3$ and constituents $\pt > 0.15$ \GeVc\ is
shown in Figure \ref{fig:UnfSpectraRAAR03}\subref{fig:UnfSpectra_R03}. 
A centrality evolution for the yield of jets is observed.
Figure \ref{fig:UnfSpectraRAAR03}\subref{fig:RAA_R03} shows the jet nuclear
modification factor $R_{AA}^{Pythia}$ for which a jet spectrum from
Pythia-Perugia\cite{Sjostrand2006} has been used as a reference. For the nuclear 
modification factor a simulated reference is used due to the limited statistics in the	\pp\ minimum bias data 
at $\sNN$=$2.76$ TeV.
A strong jet suppression is observed for central events. For more peripheral
events the suppression decreases.
This implies that the full jet energy is not captured by jets with radii $0.2$
and $0.3$ in heavy-ion events. This is also observed in the jet $R_{CP}$ as
shown in Figure \ref{fig:RCP_R03} where the jet spectrum measured in
$50$-$80$\% centrality has been used as a reference.

Figure \ref{fig:RatioR02R03CompPythia} shows that the ratio between the
measured jet spectra for radii of $R=0.2$ and $R=0.3$ is consistent with jet
production in vacuum for central and peripheral events. No significant jet
broadening between radii of $0.2$ and $0.3$ is observed in the ratio of the cross sections.
 \SubFig{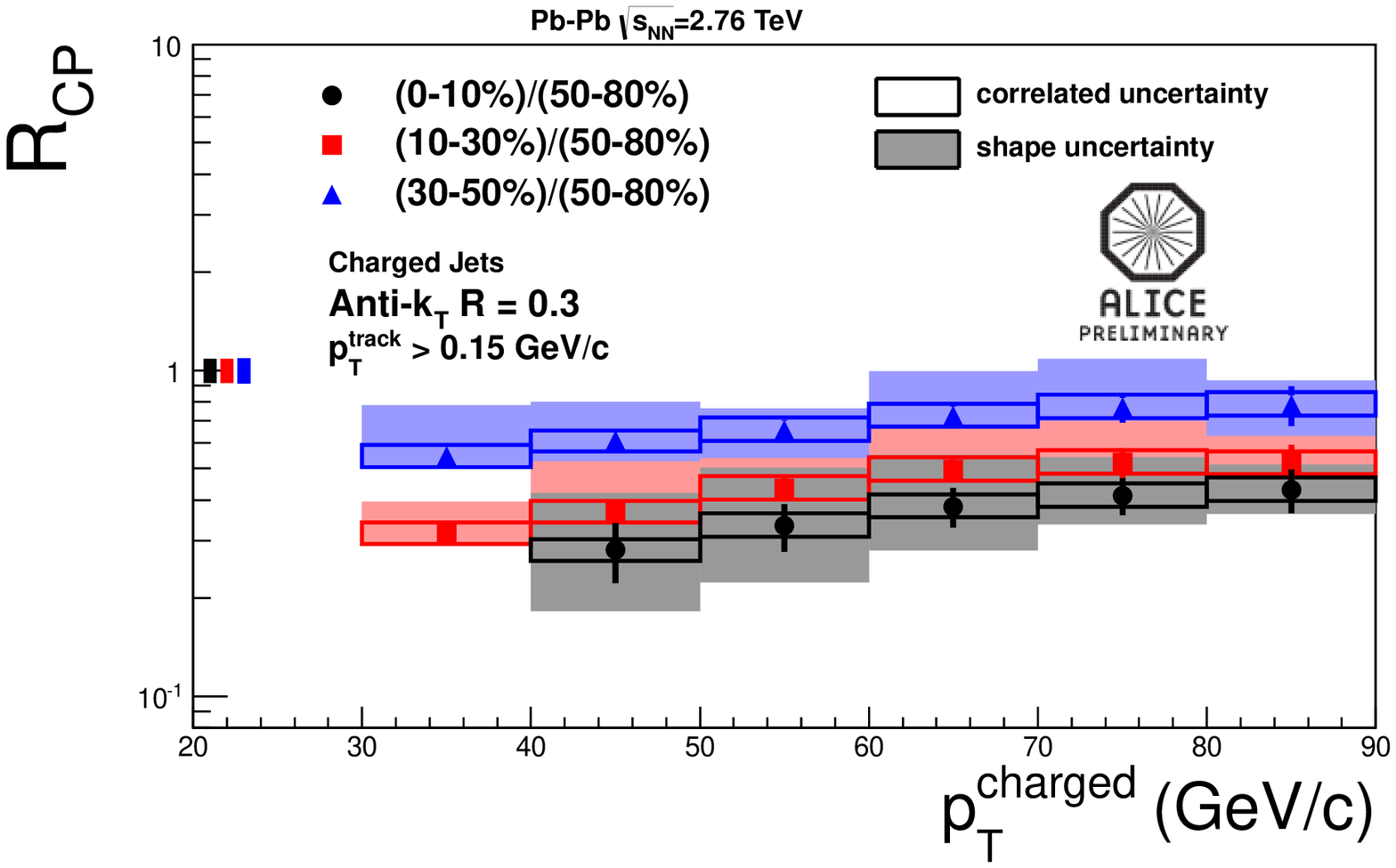}{fig:RCP_R03}{$R_{CP}$ for jet radius
   $R=0.3$. The peripheral jet spectrum corresponds to $50$-$80$\% centrality.}
        {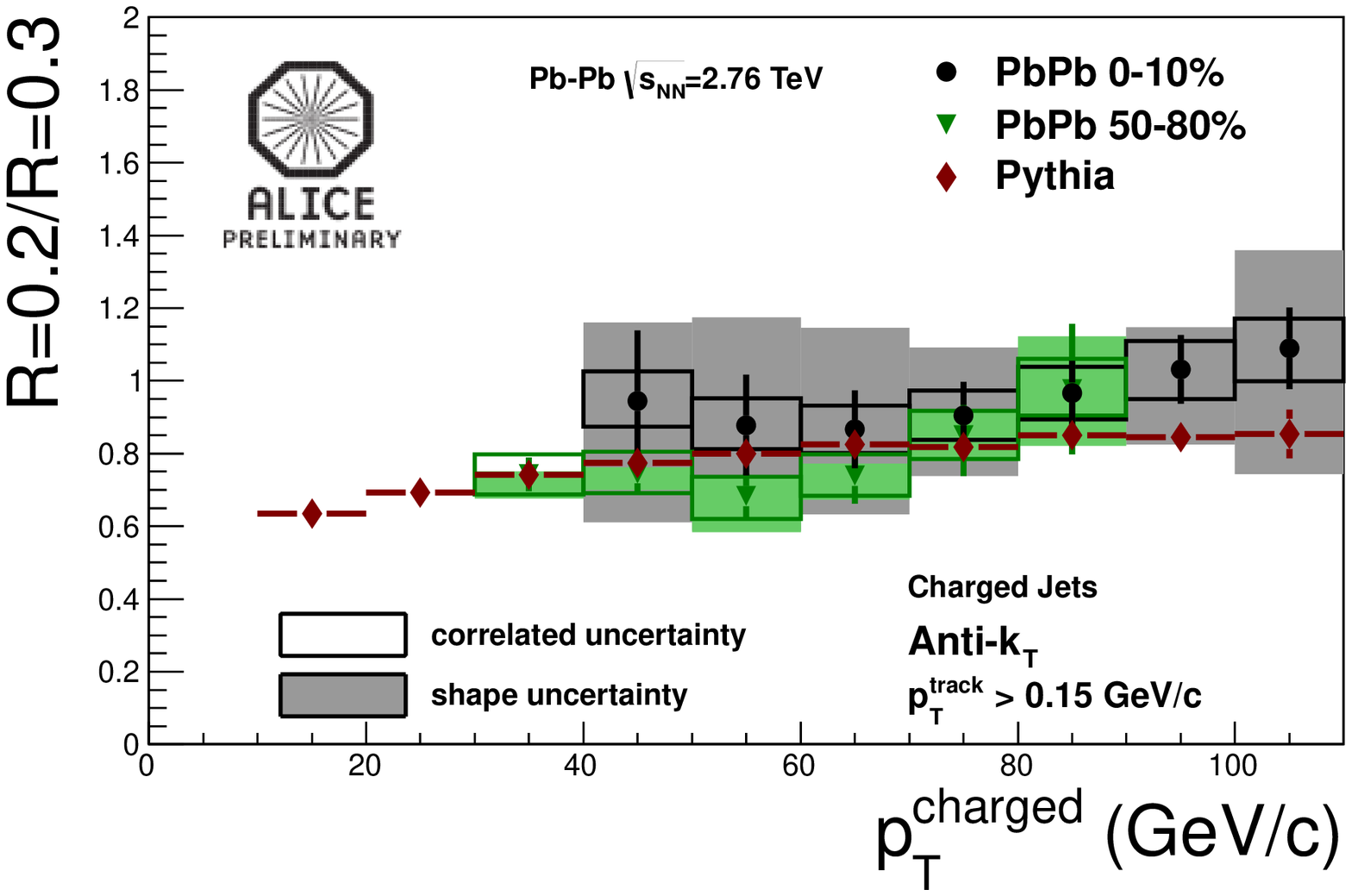}{fig:RatioR02R03CompPythia}{Ratio
          between $R=0.2$ and $R=0.3$ \PbPb jet spectra compared to
          Pythia.}  
        {Jet $R_{CP}$ and ratio between $R=0.2$ and $R=0.3$ jet spectra.}{fig:JetRCPRatioR02R03}{width=0.48\linewidth}

\begin{figure}[!h]
  \centering
  \subfigure[Comparison
  between JEWEL and measured charged jet $R_{AA}$ for jet radius $R=0.3$.]{\label{fig:RAAJEWEL_R03}
    \includegraphics[width=0.32\linewidth]{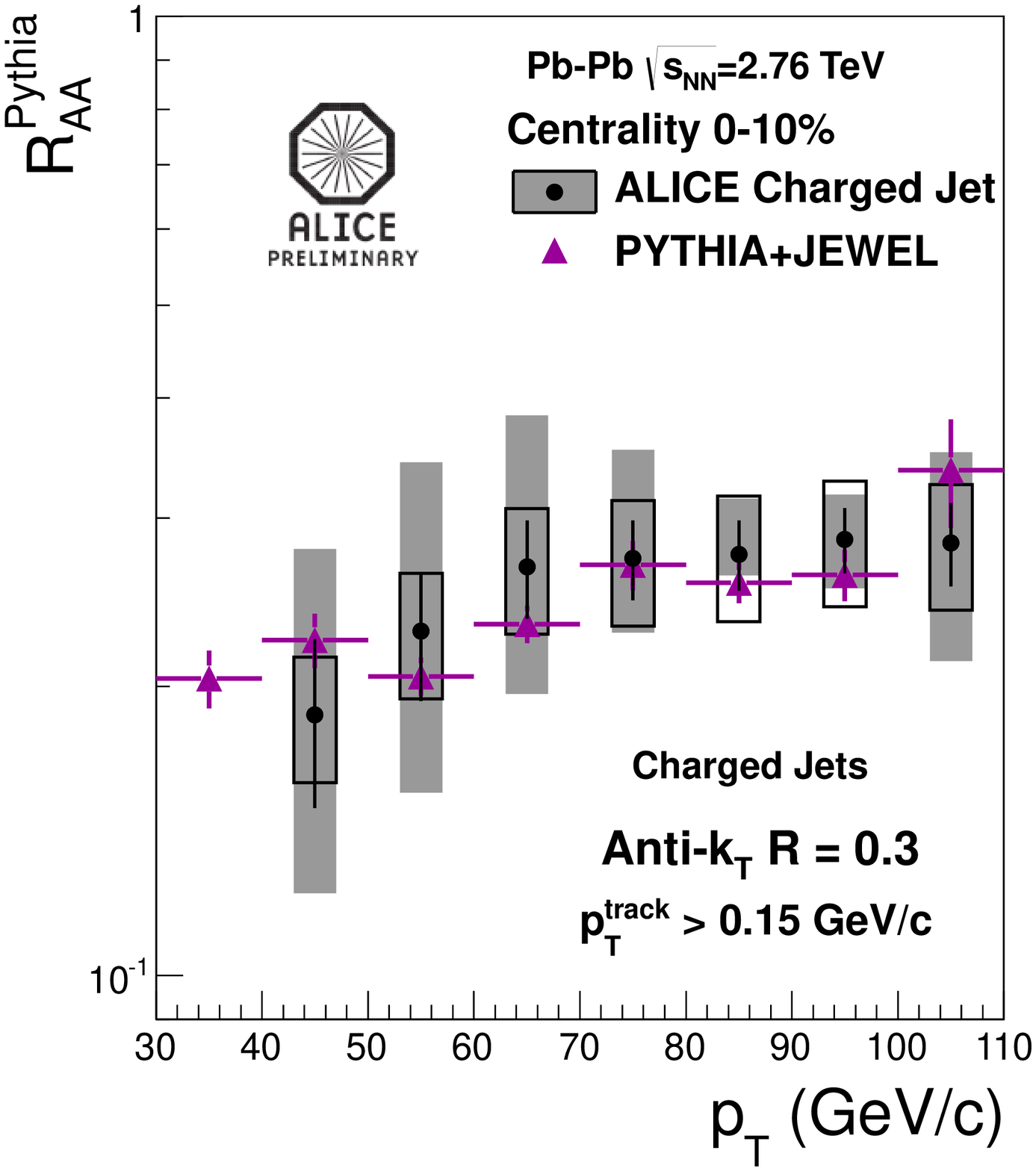}}
  \hspace{.3in}
  \subfigure[Ratio between $R=0.2$ and $R=0.3$ \PbPb jet spectra compared to
JEWEL.]{\label{fig:RatioJEWEL_R02R03}
    \includegraphics[width=0.55\linewidth]{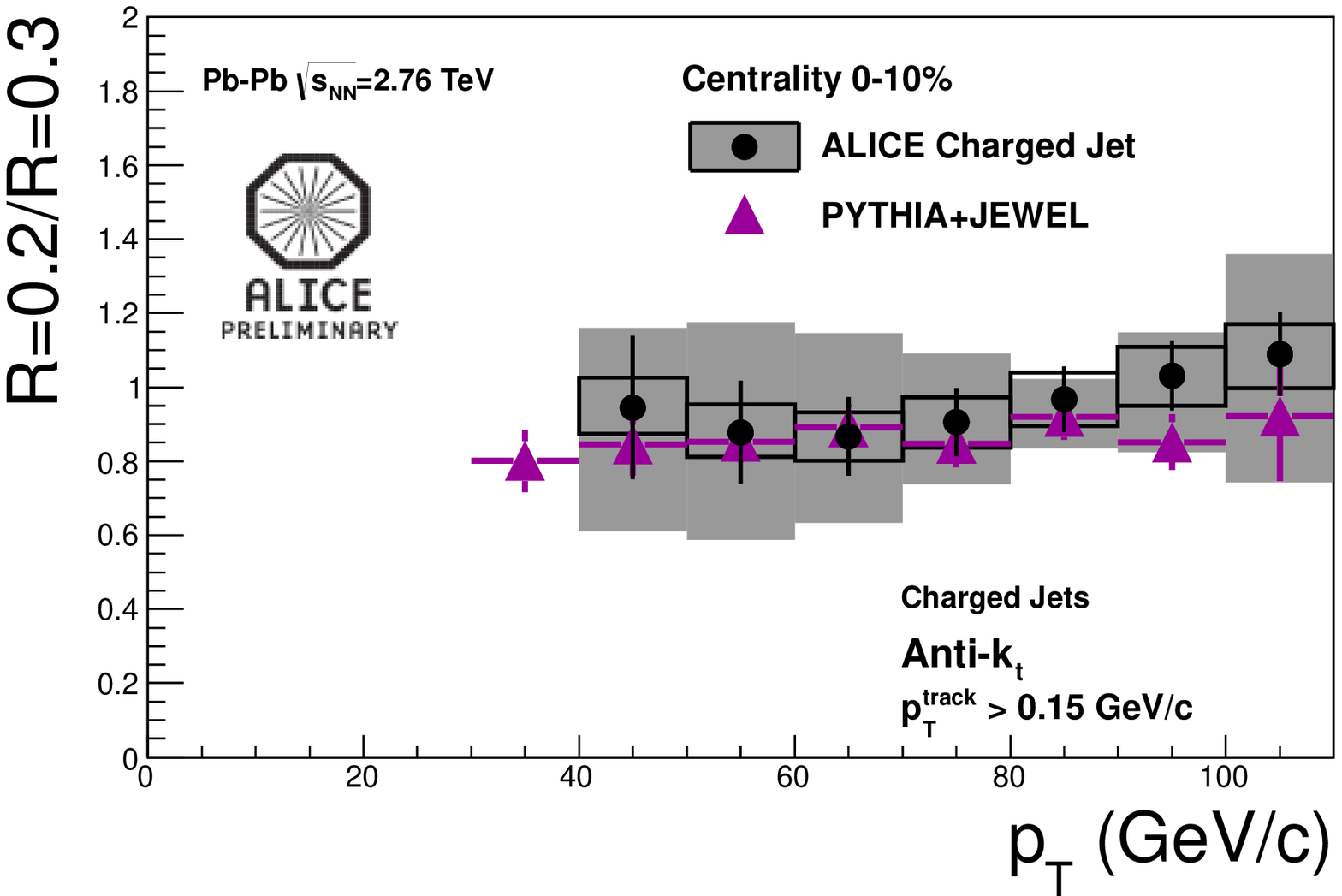}}
  \caption{Comparison of data and JEWEL energy loss MC.\label{fig:JEWEL}}
\end{figure}

The charged jet results of ALICE are compared to the JEWEL jet quenching
MC \cite{Zapp:2011ek, zapp} in Figure \ref{fig:JEWEL}. A good agreement is observed between the energy loss implementation of
JEWEL and the charged jet results from ALICE.
%





\newpage
\bibliographystyle{elsarticle-num}
\bibliography{biblio}







\end{document}